\documentclass[twocolumn,superscriptaddress]{revtex4}
\usepackage{amssymb}
\usepackage{amsthm}
\usepackage{amsmath}
\usepackage{latexsym}
\usepackage{xcolor}
\usepackage{epsfig}
\usepackage{float}
\usepackage{multirow}
\usepackage{soul}
\usepackage{xcolor}
\usepackage{ulem}

\begin{document}
\title{The shadows of quantum gravity on Bell's inequality}

\author{Hooman Moradpour}
\affiliation{Research Institute for Astronomy and Astrophysics of Maragha (RIAAM), University of Maragheh, P.O. Box 55136-553, Maragheh, Iran}
\author{Shahram Jalalzadeh}
\affiliation{Departamento de Fisica, Universidade Federal de Pernambuco, Recife, PE 50670-901, Brazil}\affiliation{Center for Theoretical Physics, Khazar University, 41 Mehseti Street, Baku, AZ1096, Azerbaijan}
\author{Hamid Tebyanian}
\affiliation{Department of Mathematics, University of York, Heslington, York, YO10 5DD, United Kingdom}

\begin{abstract}
This study delves into the validity of quantum mechanical
operators in the context of quantum gravity, recognizing the
potential need for their generalization. A primary objective is to
investigate the repercussions of these generalizations on the
inherent non-locality within quantum mechanics, as exemplified by
Bell's inequality. Additionally, the study scrutinizes the
consequences of introducing a non-zero minimal length into the
established framework of Bell's inequality. The findings
contribute significantly to our theoretical comprehension of the
intricate interplay between quantum mechanics and gravity.
Moreover, this research explores the impact of quantum gravity on
Bell's inequality and its practical applications within quantum
technologies, notably in the realms of device-independent
protocols, quantum key distribution, and quantum randomness
generation.
\end{abstract}

\maketitle
\section{Introduction}

The quantum realm is governed by the Heisenberg uncertainty
principle (HUP), which mandates that the Hamiltonian be written as
the starting point, leading to the Schrodinger equation and,
eventually, the eigenvalues and wave function of the quantum
system under consideration. In Heisenberg's formulation of quantum
mechanics (QM) in the Hilbert space, we encounter states rather
than wave functions (although they are connected). In general, QM
fails to produce satisfactory solutions for systems featuring the
Newtonian gravitational potential in their Hamiltonian. Therefore,
in conventional and widely accepted quantum mechanics, gravity is
not accounted for in terms of its operators or corresponding
Hilbert space (quantum states) carrying gravitational information.

The incompatibility of gravity and quantum mechanics is not
limited to Newtonian gravity and persists even when general
relativity is considered. On the other hand, the existence of
gravity, even in a purely Newtonian regime, leads to a non-zero
minimum (of the order of $10^{-35}\textmd{m}$ (Planck length)
\cite{Mead:1964zz}) for the uncertainty in position measurement
\cite{Mead:1966zz, Mead:1964zz, Kempf:1994su,
Hossenfelder:2012jw}. Consistently, various scenarios of quantum
gravity (QG), like String theory, also propose a non-zero minimal
for the length measurement \cite{Kempf:1994su,
Hossenfelder:2012jw}. The non-zero minimal length existence may
affect the operators, and it leads to the generalization of HUP,
called generalized uncertainty principle (GUP) \cite{Kempf:1994su,
Hossenfelder:2012jw} that becomes significant at scales close to
the Planck length and may even justify a modified gravity
\cite{Wojnar:2023bvv, Ali:2024tbd}. This concept has profound
implications for our understanding of space and time at the most
fundamental level. It seems that minimal length is not merely a
mathematical artifact of the theory but a physical reality that
could have observable consequences. This is a crucial point, as it
suggests that the effects of quantum gravity could be detected in
experiments, a distinction subtle but essential, as it affects how
we understand the physical implications of the GUP
\cite{Bosso2023}. Furthermore, understanding the effects of the
GUP on various quantum mechanical phenomena is an important issue
traced in diverse works like Refs.~\cite{Das:2008kaa} where the
GUP implications on $i$) the behavior of various oscillators,
$ii$) the transformations of space and time, and $iii$) the
emergence of a cutoff in the energy spectrum of quantum systems
have been investigated.

Operators and system states in QG may differ from those in QM.
They are, in fact, functions of ordinary operators that appear in
QM \cite{Hossenfelder:2012jw}. For instance, when considering the
first order of the GUP parameter ($\beta$), we find that the
momentum operator $\hat{P}$ can be expressed as $\hat{p}(1+\beta
\hat{p}^2)$, where $\hat{P}$ and $\hat{p}$ represent momentum
operators in QG and QM, respectively. In this representation,
$\beta$ is positive, the position operator remains unchanged
\cite{Hossenfelder:2012jw}, and GUP is written as
$\Delta\hat{x}\Delta\hat{P}\geq\frac{\hbar}{2}[1+\beta(\Delta\hat{P})^2]$.
Here, although $\beta$ seems to be a positive parameter
\cite{Kempf:1994su,Hinrichsen:1995mf,Kempf:1996ss} related to a
minimal length (of the order of the Planck length
($\equiv10^{-35}\textmd{m}$)) as
$\Delta\hat{x}=\hbar\sqrt{\beta}$, models including negative
values for $\beta$ have also been proposed \cite{Du:2022mvr}.
Current experiments and theoretical ideas predict a large range
for the upper bound of its value
\cite{Hossenfelder:2012jw,Bosso2023,Moradpour:2022lcc,Aghababaei:2022rqi,Scardigli:2019pme}.
Therefore, it follows that gravity could impact our understanding
of classical physics-based operator sets that have been
established by QM \cite{sur, sur1}. Consequently, it is possible
to write $\hat{O}=\hat{o}+\beta\hat{o}_p$ for some operators,
where $\hat{O}$ and $\hat{o}$ are operators in QG and QM,
respectively, and $\hat{o}_p$ is the first-order correction
obtained using perturbation theory \cite{Aghababaei:2022jxd}. It
should also be noted that as the position operator does not change
in the above mentioned representation \cite{Hossenfelder:2012jw},
we have $\hat{o}_p=0$ for this operator.

The discovery of quantum non-locality goes back to the famous
thought experiment by Einstein, Podolsky and Rosen (EPR) designed
to challenge the quantum mechanics \cite{Einstein:1935rr}. It
clearly shows the role of HUP in emerging quantum non-locality and
thus, an advantage for QM versus classical mechanics \cite{Hei,
Lan}. In order to establish a border between classical physics and
quantum mechanical phenomena, J. S. Bell~\cite{bell} introduces
his inequality including the maximum possible correlation in
classical physics that respects the locality. In the presence of
quantum non-locality, this inequality is violated \cite{bell2},
and the first experimental evidences of its violation (and thus,
the existence of quantum non-locality) have been reported by
Aspect et al. \cite{aspect1,aspect2,aspect3}. Interestingly
enough, the quantum non-locality is also predicted in single
particle systems \cite{sing1,sing2}.

Motivated by the correlation between HUP and quantum non-locality
(which is easily demonstrated in the square of Bell's inequality)
\cite{Lan, Hei, Einstein:1935rr}, as well as the impact of GUP on
operators, particularly angular momentum \cite{Bosso:2016frs,
Bosso:2022ogb}, recent studies have revealed that minimal length
can alter the square of Bell's operator \cite{Aghababaei:2021yzx}.
Furthermore, GUP can affect the entanglement between energy and
time, as evidenced by the results of a Franson experiment (which
serves as a testing setup for time-energy entanglement)
\cite{Aghababaei:2022rqi}. Table \ref{tab1} clearly displays the
generally expected modifications to operators and states resulting
from minimal length. The term $|\psi\rangle_{p}$ indicates an
increase in a quantum superposition, which is a probabilistic
signal for entanglement enhancement \cite{sur, sur1} and
therefore, non-locality beyond quantum mechanics
\cite{Popescu:2014wva}. It is apparent that gravity impacts the
information bound \cite{Aghababaei:2022jxd}. Indeed, studying the
effects of gravity on quantum entanglement is a long-standing
topic which will also establish ways to test the quantum aspects
of gravity. In this regard, many efforts have been made based on
the Newtonian gravity and its quantization and their effects on
the quantum entanglement
\cite{Kafri:2014hlh,Roccati:2022yqz,Christodoulou:2022knr,Liu:2023ycc,Lami:2023gmz,Marchese:2024zfu}.

\begin{table}
\centering
\begin{tabular}{ |c|c|c|c| }%{l|r}
\hline
QM & QG \\\hline
$\Delta\hat{x}\Delta\hat{p}\geq\frac{\hbar}{2}$ (HUP) & $\Delta\hat{x}\Delta\hat{P}\geq\frac{\hbar}{2}[1+\beta(\Delta\hat{P})^2]$ (GUP) \\
$\hat{o}$ & $\hat{O}=\hat{o}+\beta\hat{o}_p$ \\
$|\psi\rangle$ & $|\psi_{GUP}\rangle=|\psi\rangle+\beta|\psi\rangle_{p}$\\
\hline
%$|o\rangle^i$ & $|O\rangle_{GUP}^i=|o\rangle^i+\beta|o\rangle_{p}^i$
\end{tabular}
\caption{\label{tab1}A comparison between QM and QG (up to the
first order of $\beta$). Here, $|\psi\rangle$ and
$|\psi_{GUP}\rangle$ denote the quantum states in QM and QG,
respectively, and $|\psi\rangle_{p}$ is also calculable using the
perturbation theory. It should be noted that for operators not
affected by GUP (like the position operator in the above mentioned
representation \cite{Hossenfelder:2012jw}), there is not any
perturbation ($\hat{o}_p=0$) meaning that their corresponding
states remain unchanged.}
\end{table}

The inquiry into the influence of special and general relativity
(SR and GR, respectively) on Bell's inequality (quantum
non-locality) has been extensively studied over the years
\cite{Hacyan, vonBorzeszkowski:2000my, Gingrich:2002ota,
Peres:2002ip, Peres:2002wx}. The existing research on the effects
of SR on Bell's inequality can be classified into three general
categories, depending on the method of applying Lorentz
transformations: (i) the operators change while the states remain
unchanged, (ii) only the states undergo the Lorentz transformation
while the operators remain unaltered (the reverse of the previous
one), and (iii) both the operators and states are affected by the
Lorentz transformation \cite{Ter0, Ter1, Ter2, Ter3, Ter4, Ter5,
Kim:2004px, DV, Friis:2009va, DV1, DV2, MMM}. In order to clarify
the first two cases, consider a Lab frame, carrying a Bell state
($|\phi\rangle$) and a Bell measurement apparatus ($B$), and a
moving frame (including a Bell measurement apparatus ($B'$)) so
that they are connected to each other through the Lorentz
transformation $\Lambda$. In this manner, the moving frame faces
the Lorentz transformed Bell state $|\phi^\Lambda\rangle$, and
whenever the Lab frame looks at the Bell measurement apparatus of
the moving frame ($B'$), its Lorentz transformed is seen
($B'^\Lambda$). Now, it is apparent that using the same directions
for the Bell measurement, we find
$\langle\phi|B|\phi\rangle\neq\langle\phi^\Lambda|B'|\phi^\Lambda\rangle\neq\langle\phi|B'^\Lambda|\phi\rangle$
meaning that the maximum violation amount of Bell's inequality is
reported by both observers at different measurement directions
\cite{Ter0, Ter1, Ter2, Ter3, Ter4, Ter5, DV, Friis:2009va, DV1,
DV2, MMM}. In the third case, the moving observer is supposed to
witness a Bell measurement done in the Lab frame. The moving frame
sees $|\phi^\Lambda\rangle$ and $B^\Lambda$ (the Lorentz
transformed version of $B$) leading to
$\langle\phi|B|\phi\rangle=\langle\phi^\Lambda|B^\Lambda|\phi^\Lambda\rangle$
meaning that both the Lab observer and the moving viewer report
the same amount for the Bell measurement, simultaneously
\cite{Kim:2004px}. Furthermore, certain implications of GR and
non-inertial observers have also been addressed in
Refs.~\cite{Terashima:2003rjs, Fuentes-Schuller:2004iaz,
Alsing:2006cj, Torres-Arenas:2018vei}. Given the ongoing effort to
bridge QG with QM \cite{Ashtekar:2002sn}, exploring the effects of
QG on quantum non-locality is deemed inevitable and advantageous.

Bell's theorem suggests that certain experimental outcomes are
constrained if the universe adheres to local realism. However,
quantum entanglement, which seemingly allows distant particles to
interact instantaneously, can breach these constraints
\cite{DIBELL}. This led to cryptographic solutions like quantum
key distribution (QKD) \cite{QKD} and quantum random number
generation (QRNG) \cite{semi-DI_rusca, SDI_Avesani2022}. However,
classical noise can enter QKDs and QRNGs during implementation,
which hackers can exploit to gain partial information. A
device-independent (DI) method was developed to address this,
ensuring security when a particular correlation is detected,
irrespective of device noise. DI protocols often hinge on
non-local game violations, like the CHSH inequality \cite{DI1}.
Section \ref{app} delves into the impacts of QG on these
applications.

In this study, our primary goal is to explore the ramifications of
QG on Bell's inequality, specifically by investigating the
implications of minimal length (up to the first order of $\beta$).
To address this objective, we adopt a methodology analogous to the
three scenarios previously examined concerning the effects of SR
on quantum non-locality. To facilitate this exploration, we
categorize the existing cases into three distinct groups, which we
elaborate on in the following section. As the GUP effects become
important at energy scales close to the Planck scale, the first
case includes a quantum state produced at purely quantum
mechanical situation (low-energy) while the observer uses the Bell
measurement apparatus prepared by employing the quantum aspects of
gravity meaning that high-energy physics considerations have been
employed to build the apparatus. Therefore, we face a high-energy
affected observer (measurement), who tries to study the quantum
non-locality stored in a low-energy state (a purely quantum
mechanical state). The reversed situation is checked in the second
case, and the consequences of applying a quantum gravity-based
Bell measurement, built by considering the effects of QG, on a
state including the QG consideration are also investigated as the
third case. The paper concludes by providing a comprehensive
summary of our research findings, shedding light on the intricate
interplay between quantum mechanics and gravity, elucidating the
impact of QG on Bell's inequality, and exploring potential
applications within various quantum-based systems.

\section{Bell's inequality and the implications of QG}

In the framework of QM, assume two particles and four operators
$\hat{A}, \hat{A}^{\prime}, \hat{B}, \hat{B}^{\prime}$ with
eigenvalues $\lambda^J$ ($J\in\{\hat{A}, \hat{A}^{\prime},
\hat{B}, \hat{B}^{\prime}\}$), while the first (second) two
operators act on the first (second) particle. Now, operators
$\hat{j}=\frac{\hat{J}}{|\lambda^J|}\in\{\hat{a},
\hat{a}^{\prime}, \hat{b}, \hat{b}^{\prime}\}$ have eigenvalues
$\pm1$, and Bell's inequality is defined as
\begin{eqnarray}\label{1}
\big\langle\hat{B}\big\rangle\equiv\big\langle\hat{a}(\hat{b}+\hat{b^{\prime}})+\hat{a^{\prime}}(\hat{b}-\hat{b^{\prime}})\big\rangle\leq2.
\end{eqnarray}

Taking into account the effects of QG (up to the first order), the
operators are corrected as $\hat{J}_{GUP}=\hat{J}+\beta\hat{J}_p$
and
$\hat{j}_{GUP}=\frac{\hat{J}+\beta\hat{J}_p}{|\lambda^J_{GUP}|}$
where $\lambda^J_{GUP}$ represents the eigenvalue of
$\hat{J}_{GUP}$. Since QM should be recovered at the limit
$\beta\rightarrow0$, one may expect
$\lambda^J_{GUP}\simeq\lambda^J+\beta\lambda^J_p$. Moreover, as
the $\beta\lambda^J_p$ term is perturbative, it is reasonable to
expect $|\beta\frac{\lambda^J_p}{\lambda^J}|<<1$ leading to
$|\lambda^J+\beta\lambda^J_p|=|\lambda^J|(1+\beta\frac{\lambda^J_p}{\lambda^J})$.
Applying modifications to the states, operators, or both in
QG can result in three distinct situations. Similar
studies conducted on the effects of SR on Bell's inequality have
also revealed three cases \cite{Ter0, Ter1, Ter2, Ter3, Ter4,
Ter5, Kim:2004px, MMM}. Therefore, it is necessary to consider the
possibilities arising from these situations to understand the
implications of quantum gravitational modifications. In the
following paragraphs, we will examine these possibilities in
depth.

\subsubsection{Purely quantum mechanical entangled states in the presence of operators modified by QG}

Firstly, let us contemplate the scenario in which an entangled
state ($|\xi\rangle$) has been prepared away from the QG
influences. This implies that the objective has been accomplished
using purely quantum mechanical procedures. Furthermore, it is
assumed that an observer utilizes Bell measurements that are
constructed through the incorporation of operators containing the
QG corrections ($\hat{j}_{GUP}$). In the framework of QM, the
violation amount of inequality~(\ref{1}) depends on the directions
of Bell's measurements. Here, we have
$\hat{j}=\hat{j}_{GUP}+\beta(\frac{\lambda^J_p}{\lambda^J}\hat{j}_{GUP}-\frac{\hat{J}_p}{|\lambda^J|})$
inserted into Eq.~(\ref{1}) to reach
\begin{eqnarray}\label{2}
&&\!\!\!\!\!\!\!\!\!\big\langle\hat{B}_{GUP}\big\rangle\equiv\\&&\!\!\!\!\!\!\!\!\! \big\langle\hat{a}_{GUP}\big(\hat{b}_{GUP}+\hat{b}^{\prime}_{GUP}\big)+\hat{a}^{\prime}_{GUP}\big(\hat{b}_{GUP}-\hat{b}^{\prime}_{GUP}\big)\big\rangle\leq2\nonumber\\&&\!\!\!\!\!\!\!\!\!-\big\langle\beta^{\prime}_{a}\hat{a}_{GUP}\big(\hat{b}_{GUP}+\hat{b}^{\prime}_{GUP})+\beta^{\prime}_{a^{\prime}}\hat{a}^{\prime}_{GUP}(\hat{b}_{GUP}-\hat{b}^{\prime}_{GUP}\big)\big\rangle-\nonumber\\&&\!\!\!\!\!\!\!\!\! \big\langle\hat{a}_{GUP}\big(\beta^{\prime}_{b}\hat{b}_{GUP}+\beta^{\prime}_{b^{\prime}}\hat{b}^{\prime}_{GUP}\big)+\hat{a}^{\prime}_{GUP}\big(\beta^{\prime}_{b}\hat{b}_{GUP}-\beta^{\prime}_{b^{\prime}}\hat{b}^{\prime}_{GUP}\big)\big\rangle\nonumber\\&&\!\!\!\!\!\!\!\!\! + \beta^{\prime\prime}_{a}\big\langle\hat{A}_{GUP}\big(\hat{b}_{GUP}+\hat{b}^{\prime}_{GUP})+\hat{A}^{\prime}_{GUP}(\hat{b}_{GUP}-\hat{b}^{\prime}_{GUP}\big)\big\rangle+\nonumber\\&&\!\!\!\!\!\!\! \beta^{\prime\prime}_{b}\big\langle\hat{a}_{GUP}\big(\hat{B}_{GUP}+\hat{B}^{\prime}_{GUP}\big)+\hat{a}^{\prime}_{GUP}\big(\hat{B}_{GUP}-\hat{B}^{\prime}_{GUP}\big)\big\rangle,\nonumber
\end{eqnarray}

where $\beta^{\prime}_{j}=\beta\frac{\lambda^J_p}{\lambda_J}$,
$\beta^{\prime\prime}_{j}=\beta|\lambda_J|^{-1}$ and the last two
expressions have been written using
$\beta^{\prime\prime}_{a}=\beta^{\prime\prime}_{a^\prime}$ and
$\beta^{\prime\prime}_{b}=\beta^{\prime\prime}_{b^\prime}$. In
this manner, it is clearly seen that although the state is
unchanged, in general,
$\big\langle\hat{B}_{GUP}\big\rangle\neq\big\langle\hat{B}\big\rangle$
as the operators are affected by quantum features of gravity
\cite{Aghababaei:2021yzx, Bosso:2022ogb, Aghababaei:2022rqi}. In
studying the effects of SR on Bell's inequality, whenever the
states remain unchanged, and Lorentz transformations only affect
Bell's operator, a similar situation is also obtained \cite{Ter0,
Ter1, Ter2, Ter3, Ter4, Ter5, Kim:2004px, MMM}.

\subsubsection{Purely quantum mechanical measurements and quantum gravitational states}

Now, let us consider the situation in which the Bell apparatus is
built using purely quantum mechanical operators $j$, and the
primary entangled state carries the Planck scale information,
i.e., the quantum features of gravity. It means that the entangled
state is made using the $j_{GUP}$ operators. A similar case in
studies related to the effects of SR on Bell's inequality is the
case where the Bell measurement does not go under the Lorentz
transformation while the system state undergoes the Lorentz
transformation \cite{Ter0, Ter1, Ter2, Ter3, Ter4, Ter5,
Kim:2004px, MMM}. In this setup, we have
$|\xi_{GUP}\rangle=|\xi\rangle+\beta|\xi\rangle_{p}$ and thus
\begin{eqnarray}\label{3}
&&\!\!\!\!\!\!\!\!\!\!\!\! \big\langle\xi_{GUP}\big|\hat{B}\big|\xi_{GUP}\big\rangle\equiv\big\langle\hat{B}\big\rangle_{GUP}=\big\langle\hat{B}\big\rangle+2\beta\big\langle\xi\big|\hat{B}\big|\xi\big\rangle_{p}\nonumber\\ && \!\!\!\!\!\!\!\!\!\!\!\!\Rightarrow \big\langle\hat{B}\big\rangle_{GUP}\leq 2\big(1+\beta\big\langle\xi\big|\hat{B}\big|\xi\big\rangle_{p}\big).
\end{eqnarray}

Correspondingly, if one considers a Bell measurement apparatus
that yields $\big\langle\hat{B}\big\rangle=2\sqrt{2}$, then such
an apparatus cannot lead $\big\langle\hat{B}\big\rangle_{GUP}$ to
its maximum possible value whenever Lorentz symmetry is broken
\cite{Belich:2015yaa}.

\subsubsection{Bell's inequality in a purely quantum gravitational regime}

In deriving Bell's inequality, it is a significant step to ensure
that the operators' eigenvalues are only either $\pm1$, regardless
of their origin, whether it be from QM or QG. If both the Bell
measurement and the entangled state were prepared using the
quantum gravitational operators, then it is evident that
$\big\langle\xi_{GUP}\big|\hat{B}_{GUP}\big|\xi_{GUP}\big\rangle\leq2$.
This result indicates that, when considering the effects of QG on
both the state and the operators, Bell's inequality and the
classical regime's limit (which is $2$ in the inequality) remain
unchanged compared to the previous setups. The same outcome is
also achieved when it comes to the relationship between SR and
Bell's inequality, provided that both the system state and Bell's
measurement undergo a Lorentz transformation \cite{Kim:2004px}.

\section{Results}
This section studies QG's implications on Bell's inequality,
specifically within the contexts delineated earlier. The CHSH
inequality, a specific form of Bell's inequality, provides a
quantifiable limit on the correlations predicted by local
hidden-variable theories \cite{LHV}. A violation of the CHSH
inequality underscores the inability of such approaches to account
for the observed correlations in specific experiments with
entangled quantum systems, as predicted by quantum mechanics
\cite{DI_me}.

Now, we define the scenario where there are two parties where an
entangled pair is shared between them. The entangled state of two
qubits can be represented by the Bell state:
\begin{eqnarray}\label{4}
|\psi\rangle = \frac{1}{\sqrt{2}}(|00\rangle + |11\rangle)
\end{eqnarray}
Alice and Bob each measure their respective states. They can
choose between two measurement settings: $\hat{a},
\hat{a}^{\prime}$ for Alice and $\hat{b}, \hat{b}^{\prime}$ for
Bob. The measurement results can be either $+1$ or $-1$. The
expected value of the CHSH game using the above quantum strategy
and the Bell state is given in Eq. \ref{1}. Classically, the
maximum value of $\big\langle\hat{B}\big\rangle$ is $2$. However,
this value can reach $2\sqrt{2}$ with the quantum strategy,
violating the CHSH inequality.
\begin{figure}[h]
\includegraphics[width=\linewidth]{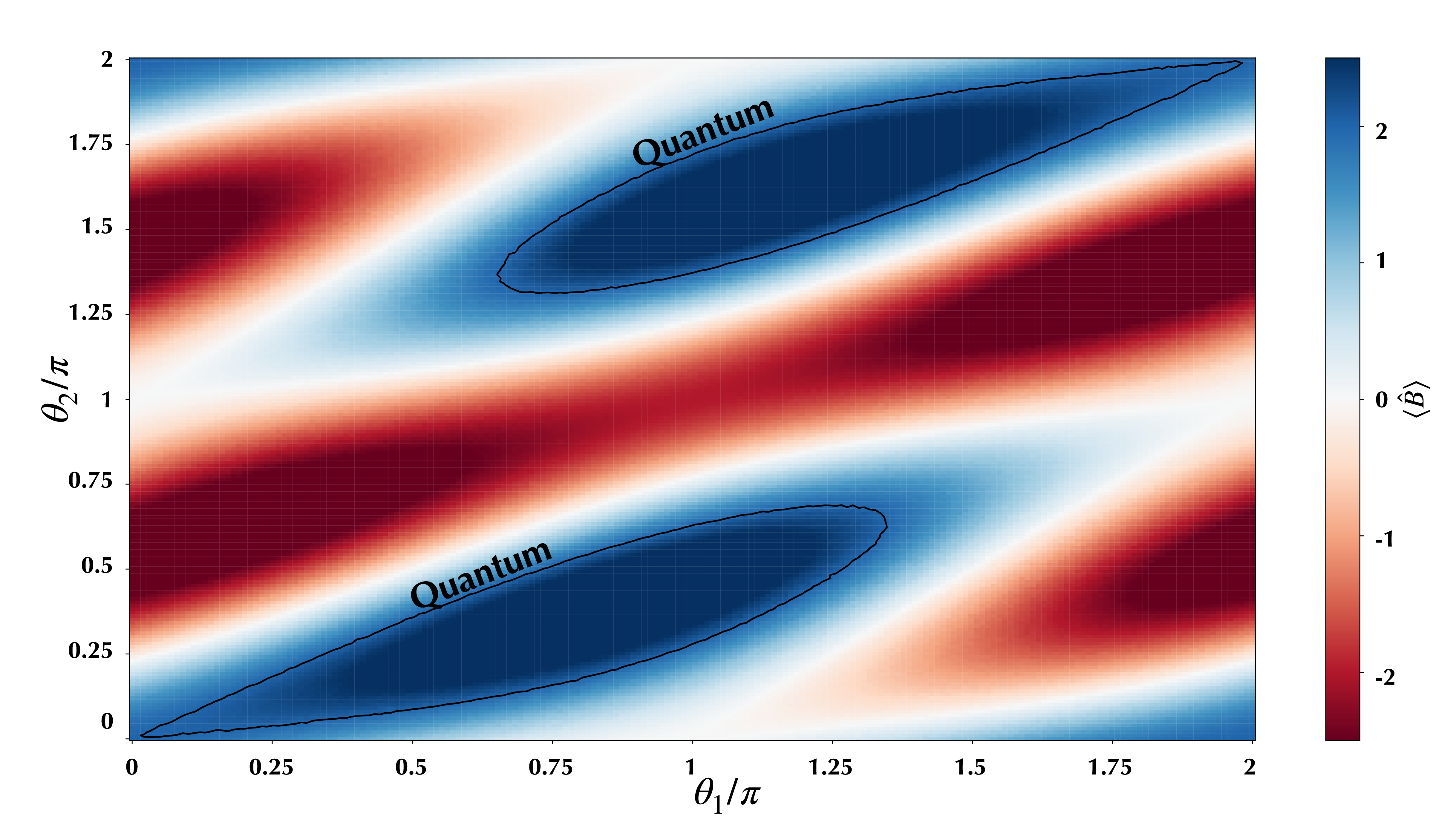}
\caption{The 2D plot of the CHSH inequality values as functions of
detection angles $\theta_1/\pi$ and $\theta_2/\pi$. Different
colors indicate different $\big\langle\hat{B}\big\rangle$ values,
with a contour distinguishing the classical and quantum regions.}
\label{chsh}
\end{figure}

Fig. \ref{chsh} illustrates that the CHSH inequality can be
surpassed by judiciously selecting the appropriate detection
angles, denoted as $\theta_1$ and $\theta_2$. The color bar
quantitatively represents the value of the inequality,
highlighting two distinct regions where the value exceeds the
classical limit of 2. In Fig. \ref{chsh}, the simulation of Bell's
inequality is conducted solely based on QM representations without
incorporating QG impact.

\begin{figure*}[ht!]
\includegraphics[width=\textwidth]{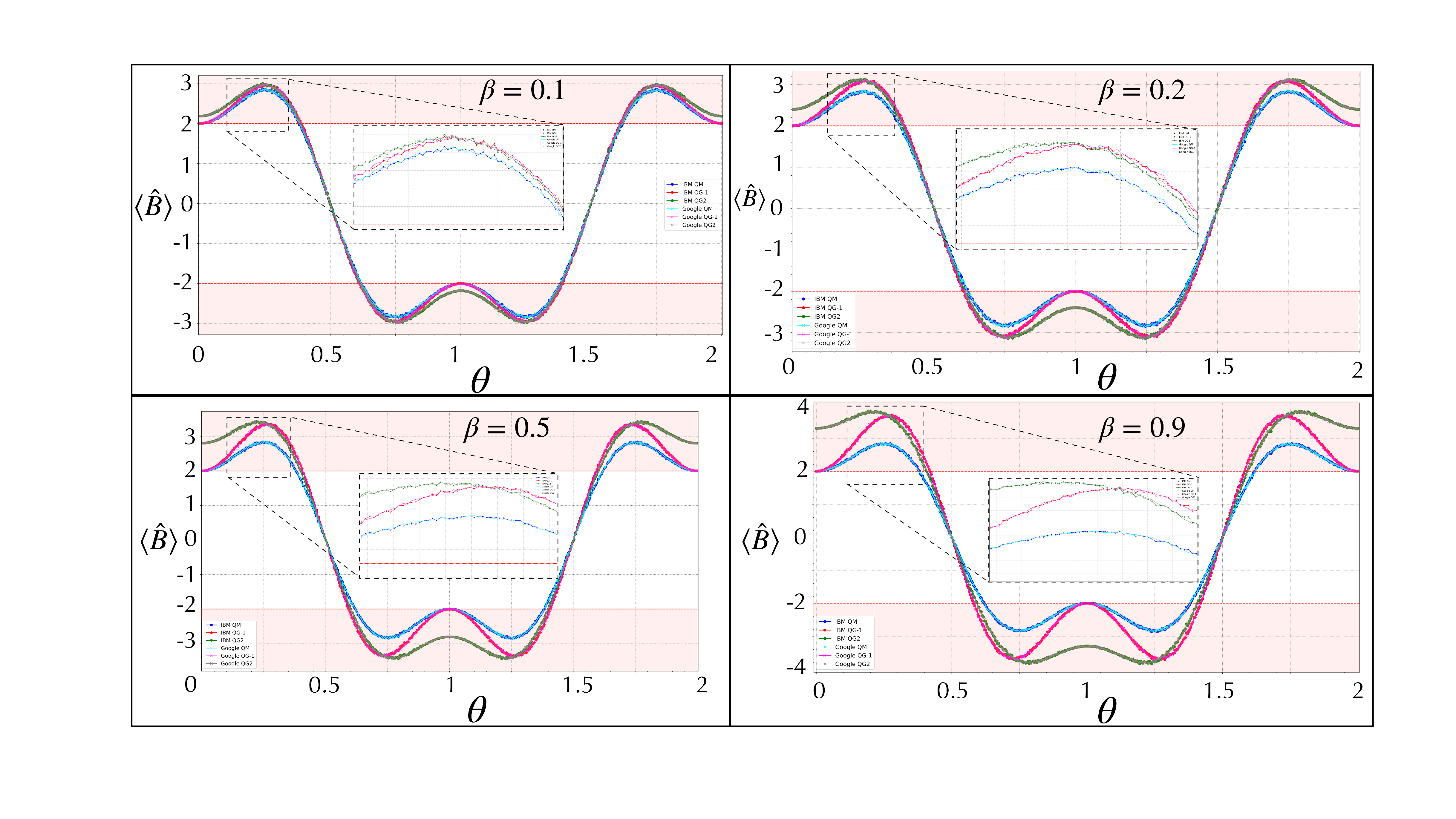}
\caption{Bell inequality values are plotted against the rotation
angle \(\theta\), illustrating the effects of varying \(\beta\)
values: 0.1, 0.2, 0.5, and 0.9. This plot comprehensively compares
six curves, each representing simulations performed on two
different quantum computing platforms: IBM and Google. For each
platform, the curves are coloured distinctly blue, red, and green
for IBM quantum computer simulations representing quantum
mechanical predictions (QM), first quantum gravitational
corrections (QG-1), and second quantum gravitational corrections
(QG-2), respectively; similarly, cyan, pink, and grey represent
the same sequence of calculations performed using a Google quantum
simulator. The remarkable overlay of curves from the two platforms
demonstrates consistent agreement, reinforcing the computational
models' reliability. An inset within the figure provides a
zoomed-in view to examine further the regions where the curves
closely approach or reach the theoretical maximum violation. This
feature is crucial for better comparing subtle differences between
the curves and understanding the implications of each model.
Notably, the maximum violation observed does not exceed the limit
of 4, consistent with the boundaries set by the Boxworld theorem.
This boundary is a crucial benchmark in general probabilistic
theories. It indicates that while the quantum mechanical
violations are significant, they do not exceed what is
theoretically possible under models that assume no
faster-than-light (superluminal) communication.}
\label{Fig::qgres} \centering
\end{figure*}

Next, we consider the QG impact on Bell's inequality for various
cases; better to say, we extend the well-known Bell inequality to
account for the effects of QG. Equations \ref{2} and \ref{3}
introduce new terms that are parameterized by $\beta$, a constant
that quantifies the strength of quantum gravitational effects.
These equations represent the modified Bell inequalities in the
presence of QG. To explore the implications of these
modifications, we plot, see Fig. \ref{Fig::qgres}, the degree of
Bell inequality violation, denoted as
$\big\langle\hat{B}\big\rangle$, as a function of $\theta$ for
various angles $\beta$. Each sub-figure in Fig. \ref{Fig::qgres}
presents six curves representing simulations conducted on two
different quantum computing platforms: IBM and Google. For IBM,
the curves are colour-coded as blue, red, and green, corresponding
to quantum mechanical predictions, first quantum gravitational
corrections, and second quantum gravitational corrections,
respectively. The Google platform uses cyan, pink, and grey to
represent the same sequence of calculations. These simulations are
repeated multiple times to account for noise, defects in quantum
computer circuits, and errors in computation and simulation using
the independent platforms of IBM and Google. The insets in each
figure show that the results from both platforms are in agreement,
confirming the reliability of the findings. The maximum violation
observed in the presence of quantum gravity remains below 4,
adhering to the theoretical limit set by the world box scenario
\cite{BOX}.

The results notably indicate an escalating violation of the Bell
inequality with the introduction of QG. As the parameter $\beta$
increases, the violation surpasses the quantum mechanical limit of
$ \sqrt{8} $, signifying a more pronounced breach of the
inequality. This implies that the presence of quantum
gravitational effects could lead to a more pronounced violation of
the Bell inequality than what is predicted by standard quantum
mechanics.

\section{Applications}
\label{app} QKD and QRNG represent two extensively researched and
commercially implemented areas where the applications of quantum
mechanics come to life. While quantum mechanics underpins the
security of these systems, experimental imperfections can
introduce vulnerabilities. To address this, DI protocols have been
developed. These protocols harness the non-local correlations
inherent in quantum entanglement. Importantly, they do not rely on
an intricate understanding of the devices in use; their security
is grounded solely in the observed violation of non-local
correlations, such as the Bell inequalities. This approach offers
a robust solution to the security challenges posed by device
imperfections \cite{DI2, DI3}.

In DI QKD, two distant parties share an entangled quantum state.
They perform measurements on their respective parts of the state,
and due to the non-local nature of entanglement, the outcomes of
these measurements are correlated in a way that disobeys classical
explanation. These correlations serve as the foundation for key
generation, with the security of the key guaranteed by the
violation of Bell inequalities. Basically, any eavesdropper
attempting to intercept or tamper with the quantum states would
disrupt these correlations, making their presence detectable.

The security and randomness of DI QRNG do not depend on trusting
the intrinsic workings of the devices. Traditional QRNGs require
detailed models and assumptions about the device, but in DI QRNGs,
as long as observed outcomes violate Bell inequalities, one can be
assured of the randomness. With the rise of quantum computers,
many cryptographic methods are at risk. Nevertheless, the
unpredictability in DI QRNG is more than just computationally hard
for quantum computers; it's theoretically impossible to predict
due to the inherent randomness of quantum processes
\cite{semi-DI_rusca, SDI_Avesani2022}.

Incorporating the effects of QG in quantum information science and
technology becomes an intellectual exercise and a practical
necessity. Given the results in the previous section that QG
effects can significantly enhance the violation of Bell
inequalities, let us consider its implications for quantum
information science and technology and its applications.

The security of QKD is guaranteed by the quantum mechanical
violation of Bell inequalities; increasing the violation value of
Bell's inequality makes QKD even more secure against attacks. This
disturbance changes the quantum correlations between Alice's and
Bob's measurements. In other words, if the eavesdropper is
listening in, the observed violations of Bell's inequalities at
Alice's and Bob's ends will reduce, moving closer to what would be
expected classically. Thus, if you start with a higher violation
of Bell's inequalities (thanks to QG effects), you are raising the
"quantumness" of your initial state. The higher this initial
level, the more sensitive your system becomes to any eavesdropping
activities. A significant drop in the observed Bell inequality
violation from this higher baseline would more quickly and
definitively signal the presence of eavesdropping, thus enabling
quicker and more reliable detection of any security breaches.

DI protocols prevent the need for trust in the hardware by
utilizing Bell inequality violations the greater the violation,
the higher the level of security. The introduction of QG effects
adds an additional layer of robustness to DI protocols, fortifying
them through quantum mechanical principles and integrating
fundamental theories of nature. Similarly, for QRNGs, a heightened
violation signifies a more quantum-coherent system, enhancing the
quality of randomness, which comprises not merely an incremental
advancement but a paradigmatic leap in the entropy of the
generated random numbers. Consequently, this reduces the
computational time required to achieve a given level of randomness
and unpredictability, analogous to transitioning from conventional
vehicular propulsion to advanced warp drives, all while adhering
to the fundamental constraints of space-time.

More importantly, quantum gravity could offer richer quantum
correlations in multipartite systems. Imagine a quantum network
secured by quantum gravity effects each additional party would
enhance not just the computational power but the security,
generating what could be termed "quantum gravity-secured
entanglement." Enabling a brand-new platform for multiparty
quantum computations and secret sharing protocols.

In summary, enhanced violations of Bell inequalities render QKD
virtually impregnable, elevate QRNGs to sources of high-entropy
randomness, and establish DI protocols as the epitome of
trust-free security mechanisms. Dismissing QG as a purely academic
endeavor could overlook its potential as a critical element in
safeguarding quantum data against even the most advanced
computational threats. If quantum mechanics is considered the apex
of security and efficiency, the advent of QG compels a
reevaluation. It promises to redefine the boundaries of what is
secure, efficient, and trustworthy in quantum technologies.
\section{conclusion}

The study can be summarized by its two main components: $i$) the
origin of entangled states and $ii$) Bell's measurement.
Furthermore, the study has introduced the possibility of three
outcomes depending on which cornerstone carries the quantum
gravitational modifications. The first two scenarios suggest that
if only one of the foundations stores the effects of QG, then a
precise Bell measurement (depending on the value of $\beta$) could
detect the effects of QG. This is due to the differences between
$\big\langle\hat{B}\big\rangle$,
$\big\langle\hat{B}_{GUP}\big\rangle$, and
$\big\langle\hat{B}\big\rangle_{GUP}$. In the third case, Bell's
inequality remains invariant if we consider the quantum aspects of
gravity on both the states and the operators. Moreover, the
results demonstrate that the presence of QG enhances Bell's
inequality violation, thereby offering avenues for improving the
security and performance of  DI QRNG and QKD protocols.

\section*{Acknowledgement}
S.J. acknowledges financial support from the National Council for
Scientific and Technological Development--CNPq, Grant no.
308131/2022-3. H. T. acknowledge the Quantum Communications Hub of the UK Engineering and Physical Sciences Research Council (EPSRC) (Grant Nos. EP/M013472/1 and EP/T001011/1).


\begin{thebibliography}{99}

\bibitem{Mead:1964zz}
C.~A.~Mead,
``Possible Connection Between Gravitation and Fundamental Length,''
Phys. Rev. \textbf{135}, B849-B862 (1964)
doi:10.1103/PhysRev.135.B849

\bibitem{Mead:1966zz}
C.~A.~Mead,
``Observable Consequences of Fundamental-Length Hypotheses,''
Phys. Rev. \textbf{143}, 990-1005 (1966)
doi:10.1103/PhysRev.143.990

\bibitem{Kempf:1994su}
A.~Kempf, G.~Mangano and R.~B.~Mann,
``Hilbert space representation of the minimal length uncertainty relation,''
Phys. Rev. D \textbf{52}, 1108-1118 (1995)
doi:10.1103/PhysRevD.52.1108
[arXiv:hep-th/9412167 [hep-th]].

\bibitem{Hossenfelder:2012jw}
S.~Hossenfelder,
``Minimal Length Scale Scenarios for Quantum Gravity,''
Living Rev. Rel. \textbf{16}, 2 (2013)
doi:10.12942/lrr-2013-2
[arXiv:1203.6191 [gr-qc]].

\bibitem{Wojnar:2023bvv}
A.~Wojnar, ``Unveiling phase space modifications: A clash of
modified gravity and the generalized uncertainty principle,''
Phys. Rev. D \textbf{109}, no.2, 024011 (2024)
doi:10.1103/PhysRevD.109.024011 [arXiv:2311.14066 [gr-qc]].
\bibitem{Ali:2024tbd}
A.~F.~Ali and A.~Wojnar, ``A covariant tapestry of linear GUP,
metric-affine gravity, their Poincar\'e algebra and entropy
bound,'' Class. Quant. Grav. \textbf{41}, no.10, 105001 (2024)
doi:10.1088/1361-6382/ad3ac7 [arXiv:2401.05941 [gr-qc]].
\bibitem{Bosso2023} P.~Bosso, G.~G.~Luciano,
L.~Petruzziello and F.~Wagner, ``30 years in: Quo vadis
generalized uncertainty principle?,'' Class. Quant. Grav.
\textbf{40}, no.19, 195014 (2023) doi:10.1088/1361-6382/acf021
[arXiv:2305.16193 [gr-qc]].
\bibitem{Das:2008kaa} S.~Das and E.~C.~Vagenas,
``Universality of Quantum Gravity Corrections,'' Phys. Rev. Lett.
\textbf{101}, 221301 (2008) doi:10.1103/PhysRevLett.101.221301
[arXiv:0810.5333 [hep-th]].
\bibitem{Hinrichsen:1995mf}
H.~Hinrichsen and A.~Kempf, ``Maximal localization in the presence
of minimal uncertainties in positions and momenta,'' J. Math.
Phys. \textbf{37}, 2121-2137 (1996) doi:10.1063/1.531501
[arXiv:hep-th/9510144 [hep-th]].
\bibitem{Kempf:1996ss}
A.~Kempf, ``On quantum field theory with nonzero minimal
uncertainties in positions and momenta,'' J. Math. Phys.
\textbf{38}, 1347-1372 (1997) doi:10.1063/1.531814
[arXiv:hep-th/9602085 [hep-th]].
\bibitem{Du:2022mvr}
X.~D.~Du and C.~Y.~Long, ``New generalized uncertainty principle
with parameter adaptability for the minimum length,'' JHEP
\textbf{10}, 063 (2022) doi:10.1007/JHEP10(2022)063
[arXiv:2208.12918 [gr-qc]].
\bibitem{Moradpour:2022lcc}
H.~Moradpour, A.~H.~Ziaie and N.~Sadeghnezhad, ``Generalized
uncertainty principle and burning stars,'' Front. Astron. Space
Sci. \textbf{9}, 936352 (2022) doi:10.3389/fspas.2022.936352
[arXiv:2206.13940 [gr-qc]].
\bibitem{Scardigli:2019pme}
F.~Scardigli, ``The deformation parameter of the generalized
uncertainty principle,'' J. Phys. Conf. Ser. \textbf{1275}, no.1,
012004 (2019) doi:10.1088/1742-6596/1275/1/012004
[arXiv:1905.00287 [hep-th]].
\bibitem{Aghababaei:2022rqi}
S.~Aghababaei, H.~Moradpour, ``Generalized uncertainty principle
and quantum non-locality,'' Quant. Inf. Proc. \textbf{22}, no.4,
173 (2023) doi:10.1007/s11128-023-03920-7 [arXiv:2202.07489
[quant-ph]].

\bibitem{sur}
``Survey the foundations,'' Nat. Phys. 18, 961 (2022). https://doi.org/10.1038/s41567-022-01766-x

\bibitem{sur1}
J. R. Hance, S. Hossenfelder, ``Bell's theorem allows local
theories of quantum mechanics,'' Nat. Phys. 18, 1382 (2022).
https://doi.org/10.1038/s41567-022-01831-5

\bibitem{Aghababaei:2022jxd}
S.~Aghababaei, H.~Moradpour, S.~S.~Wani, F.~Marino, N.~A.~Shah and
M.~Faizal. ``Effective information bounds in modified quantum
mechanics,'' Eur. Phys. J. C 84, 404 (2024)
https://doi.org/10.1140/epjc/s10052-024-12749-y [arXiv:2211.09227
[quant-ph]].

\bibitem{Einstein:1935rr}
A.~Einstein, B.~Podolsky and N.~Rosen, ``Can quantum mechanical
description of physical reality be considered complete?,'' Phys.
Rev. \textbf{47}, 777-780 (1935) doi:10.1103/PhysRev.47.777

\bibitem{Hei}
W.~Heisenberg, ``Uber den anschaulichen Inhalt der
quantentheoretischen Kinematik und Mechanik,'' Z. Physik 43,
172-198 (1927). doi:10.1007/BF01397280

\bibitem{Lan}
L.~Landau, ``Experimental tests of general quantum theories,``
Letters in Mathematical Physics 14, 33-40 (1987)
doi:10.1007/BF00403467

\bibitem{bell} J. S. Bell, ``On the Einstein Podolsky
Rosen paradox,'' Physics (N.Y). \textbf{1}, 195 (1964).
\bibitem{bell2} N. Brunner, D. Cavalcanti, S. Pironio, V. Scarani, and S. Wehner, ``Bell nonlocality,'' Rev. Mod. Phys. \textbf{86}, 419--478 (2014).
\bibitem{aspect1} A. Aspect, P. Grangier, and G. Roger, ``Experimental tests of realistic local theories via Bell's theorem,'' Phys. Rev. Lett. \textbf{47}, 460--463 (1981).
\bibitem{aspect2} A. Aspect, P. Grangier, and G. Roger, ``Experimental realization of Einstein-Podolsky-Rosen-Bohm Gedankenexperiment: a new violation of Bell's inequalities,'' Phys. Rev. Lett. \textbf{49}, 91--94 (1982).
\bibitem{aspect3} A. Aspect, P. Grangier, and G. Roger, ``Experimental test of Bell's inequalities using time-varying analyzers,'' Phys. Rev. Lett. \textbf{49}, 1804--1807 (1982).
\bibitem{sing1} J. A. Dunningham, V. Vedral, ``Nonlocality of a single particle,'' Phys. Rev. Lett. \textbf{99}, 180404 (2007).
\bibitem{sing2} J. J. Cooper, J. A. Dunningham, ``Single particle nonlocality with completely independent reference states,'' New J. Phys. \textbf{10}, 113024
(2008).


\bibitem{Bosso:2016frs}
P.~Bosso and S.~Das,
``Generalized Uncertainty Principle and Angular Momentum,''
Annals Phys. \textbf{383}, 416-438 (2017)
doi:10.1016/j.aop.2017.06.003
[arXiv:1607.01083 [gr-qc]].

\bibitem{Bosso:2022ogb}
P.~Bosso, L.~Petruzziello, F.~Wagner and F.~Illuminati,
``Spin operator, Bell nonlocality and Tsirelson bound in quantum-gravity induced minimal-length quantum mechanics,''
Commun. Phys. \textbf{6}, no.1, 114 (2023)
doi:10.1038/s42005-023-01229-6
[arXiv:2207.10418 [quant-ph]].

\bibitem{Aghababaei:2021yzx}
S.~Aghababaei, H.~Moradpour and H.~Shabani,
``Quantum gravity and the square of Bell operators,''
Quant. Inf. Proc. \textbf{21}, no.2, 57 (2022)
doi:10.1007/s11128-021-03397-2
[arXiv:2106.14400 [quant-ph]].

\bibitem{Popescu:2014wva}
S.~Popescu,
``Nonlocality beyond quantum mechanics,''
Nature Phys. \textbf{10}, no.4, 264-270 (2014)
doi:10.1038/nphys2916

\bibitem{Kafri:2014hlh} D.~Kafri, G.~J.~Milburn and
J.~M.~Taylor, ``Bounds on quantum communication via Newtonian
gravity,'' New J. Phys. \textbf{17}, no.1, 015006 (2015)
doi:10.1088/1367-2630/17/1/015006 [arXiv:1404.3214 [quant-ph]].
\bibitem{Roccati:2022yqz}
F.~Roccati, B.~Militello, E.~Fiordilino, R.~Iaria, L.~Burderi,
T.~Di Salvo and F.~Ciccarello, ``Quantum correlations beyond
entanglement in a classical-channel model of gravity,'' Sci. Rep.
\textbf{12}, no.1, 17641 (2022) doi:10.1038/s41598-022-22212-1
[arXiv:2205.15333 [quant-ph]].
\bibitem{Christodoulou:2022knr}
M.~Christodoulou, A.~Di Biagio, R.~Howl and C.~Rovelli, ``Gravity
entanglement, quantum reference systems, degrees of freedom,''
Class. Quant. Grav. \textbf{40}, no.4, 047001 (2023)
doi:10.1088/1361-6382/acb0aa [arXiv:2207.03138 [quant-ph]].
\bibitem{Liu:2023ycc}
S.~Liu, L.~Chen and M.~Liang, ``Multiqubit entanglement due to
quantum gravity,'' Phys. Lett. A \textbf{493}, 129273 (2024)
doi:10.1016/j.physleta.2023.129273 [arXiv:2301.05437 [quant-ph]].
\bibitem{Lami:2023gmz}
L.~Lami, J.~S.~Pedernales and M.~B.~Plenio, ``Testing the quantum
nature of gravity without entanglement,'' [arXiv:2302.03075
[quant-ph]].
\bibitem{Marchese:2024zfu}
M.~M.~Marchese, M.~Pl\'avala, M.~Kleinmann and S.~Nimmrichter,
``Newton's laws of motion can generate gravity-mediated
entanglement,'' [arXiv:2401.07832 [quant-ph]].

\bibitem{Hacyan}
S.~Hacyan,
``Relativistic invariance of Bell's inequality,''
Phys. Lett. A \textbf{288}, 59-61 (2001)
ISSN 0375-9601,
doi:10.1016/S0375-9601(01)00519-9.

\bibitem{vonBorzeszkowski:2000my}
H.~von Borzeszkowski and M.~B.~Mensky,
``EPR effect in gravitational field: Nature of nonlocality,''
Phys. Lett. A \textbf{269}, 197-203 (2000)
doi:10.1016/S0375-9601(00)00230-9
[arXiv:quant-ph/0007085 [quant-ph]].

\bibitem{Gingrich:2002ota}
R.~M.~Gingrich and C.~Adami,
``Quantum Entanglement of Moving Bodies,''
Phys. Rev. Lett. \textbf{89}, 270402 (2002)
doi:10.1103/PhysRevLett.89.270402
[arXiv:quant-ph/0205179 [quant-ph]].

\bibitem{Peres:2002ip}
A.~Peres, P.~F.~Scudo and D.~R.~Terno,
``Quantum entropy and special relativity,''
Phys. Rev. Lett. \textbf{88}, 230402 (2002)
doi:10.1103/PhysRevLett.88.230402
[arXiv:quant-ph/0203033 [quant-ph]].

\bibitem{Peres:2002wx}
A.~Peres and D.~R.~Terno,
``Quantum information and relativity theory,''
Rev. Mod. Phys. \textbf{76}, 93-123 (2004)
doi:10.1103/RevModPhys.76.93
[arXiv:quant-ph/0212023 [quant-ph]].

\bibitem{Ter0} P. M. Alsing and G. J. Milburn, ``On entanglement and Lorentz transformations,'' Quant. Inf. Comput. {\bf2}, 487 (2002).

\bibitem{Ter1}
H. Terashima and M. Ueda, ``Einstein-Podolsky-Rosen correlation
seen from moving observers,'' Quantum Inf. Comput. 3, 224-228
(2003) doi:10.48550/arXiv.quant-ph/0204138

\bibitem{Ter2} H. Terashima and M. Ueda, ``Relativistic  Einstein-Podolsky-Rosen correlation and Bell's inequality,'' Int. J. Quant. Inf. {\bf1}, 93 (2003).
doi:10.1142/S0219749903000061

\bibitem{Ter3} D. Ahn, H-J, Lee, Y. H. Moon and S. W. Hwang, ``Relativistic entanglement and Bells inequality,'' Phys. Rev. A {\bf67}, 012103 (2003).

\bibitem{Ter4}
D. Lee and E. Chang-Young, ``Quantum entanglement under Lorentz boost''. New J. Phys. {\bf6}, 67 (2004).

\bibitem{Kim:2004px}
W.~T.~Kim and E.~J.~Son,
``Lorentz invariant Bell's inequality,''
Phys. Rev. A \textbf{71}, 014102 (2005)
doi:10.1103/PhysRevA.71.014102
[arXiv:quant-ph/0408127 [quant-ph]].

\bibitem{Ter5}
T.~F.~Jordan, A.~Shaji and E.~C.~G.~Sudarshan,
``Lorentz transformations that entangle spins and entangle momenta,''
Phys. Rev. A \textbf{75}, 022101 (2007)
doi:10.1103/PhysRevA.75.022101
[arXiv:quant-ph/0608061 [quant-ph]].

\bibitem{DV}
J. Dunningham and V. Vedral, ``Entanglement and nonlocality of a single relativistic particle,'' Phys. Rev. A 80,
044302 (2009).

\bibitem{Friis:2009va}
N.~Friis, R.~A.~Bertlmann, M.~Huber and B.~C.~Hiesmayr,
``Relativistic entanglement of two massive particles,''
Phys. Rev. A \textbf{81}, 042114 (2010)
doi:10.1103/PhysRevA.81.042114
[arXiv:0912.4863 [quant-ph]].

\bibitem{DV1}
 P. L. Saldanha and V. Vedral, Phys. Rev. A 85, 062101
(2012).

\bibitem{DV2}
P. L. Saldanha and V. Vedral, New J. Phys. 14, 023041
(2012).

\bibitem{MMM}
H. Moradpour, S. Maghool, and S. A. Moosavi,
``Three-particle Bell-like inequalities under Lorentz transformations,''
Quantum Inf. Process. 14 (2015) 3913
doi:10.1007/s11128-015-1064-3

\bibitem{Terashima:2003rjs}
H.~Terashima and M.~Ueda,
``Einstein-Podolsky-Rosen correlation in a gravitational field,''
Phys. Rev. A \textbf{69}, 032113 (2004)
doi:10.1103/PhysRevA.69.032113
[arXiv:quant-ph/0307114 [quant-ph]].

\bibitem{Fuentes-Schuller:2004iaz}
I.~Fuentes-Schuller and R.~B.~Mann,
``Alice falls into a black hole: Entanglement in non-inertial frames,''
Phys. Rev. Lett. \textbf{95}, 120404 (2005)
doi:10.1103/PhysRevLett.95.120404
[arXiv:quant-ph/0410172 [quant-ph]].

\bibitem{Alsing:2006cj}
P.~M.~Alsing, I.~Fuentes-Schuller, R.~B.~Mann and T.~E.~Tessier,
``Entanglement of Dirac fields in non-inertial frames,''
Phys. Rev. A \textbf{74}, 032326 (2006)
doi:10.1103/PhysRevA.74.032326
[arXiv:quant-ph/0603269 [quant-ph]].

\bibitem{Torres-Arenas:2018vei}
A.~J.~Torres-Arenas, E.~O.~Lopez-Zuniga, J.~A.~Saldana-Herrera, Q.~Dong, G.~H.~Sun and S.~H.~Dong,
``Entanglement measures of W-state in noninertial frames,''
Phys. Lett. B \textbf{789}, 93-105 (2019)
doi:10.1016/j.physletb.2018.12.010
[arXiv:1810.03951 [quant-ph]].

\bibitem{Ashtekar:2002sn}
A.~Ashtekar, S.~Fairhurst and J.~L.~Willis,
``Quantum gravity, shadow states, and quantum mechanics,''
Class. Quant. Grav. \textbf{20}, 1031-1062 (2003)
doi:10.1088/0264-9381/20/6/302
[arXiv:gr-qc/0207106 [gr-qc]].

\bibitem{Belich:2015yaa}
H.~Belich, C.~Furtado and K.~Bakke,
``Lorentz symmetry breaking effects on relativistic EPR correlations,''
Eur. Phys. J. C \textbf{75}, no.9, 410 (2015)
doi:10.1140/epjc/s10052-015-3640-1
[arXiv:1508.04662 [hep-th]].

\bibitem{DI1}
R.~Colbeck and A.~Kent,
``Private randomness expansion with untrusted devices,''
Journal of Physics A: Mathematical and Theoretical \textbf{44}, 095305 (2011)
doi:10.1088/1751-8113/44/9/095305
[url:https://iopscience.iop.org/article/10.1088/1751-8113/44/9/095305]

\bibitem{DI2}
W.-Z.~Liu, M.-H.~Li, S.~Ragy, S.-R.~Zhao, B.~Bai, Y.~Liu, P. J.~Brown, J.~Zhang, R.~Colbeck, J.~Fan, and others,
``Device-independent randomness expansion against quantum side information,''
arXiv preprint arXiv:1912.11159 (2019)
[url:https://arxiv.org/abs/1912.11159]


\bibitem{semi-DI_new}
H.~Tebyanian, M.~Zahidy, M.~Avesani, A.~Stanco, P.~Villoresi, and G.~Vallone,
``Semi-device independent randomness generation based on quantum state's indistinguishability,''
Quantum Science and Technology \textbf{2021},
doi:10.1088/2058-9565/ac2047
[url:https://doi.org/10.1088/2058-9565/ac2047]
\bibitem{DI3}
R.~Colbeck and A.~Kent,
``Private randomness expansion with untrusted devices,''
Journal of Physics A: Mathematical and Theoretical \textbf{44}, 095305 (2011)
doi:10.1088/1751-8113/44/9/095305
[url:https://iopscience.iop.org/article/10.1088/1751-8113/44/9/095305]

\bibitem{semi-DI_rusca}
D.~Rusca, H.~Tebyanian, A.~Martin, and H.~Zbinden,
``Fast self-testing quantum random number generator based on homodyne detection,''
Applied Physics Letters \textbf{116}, 264004 (2020)
doi:10.1063/5.0011479
[url:https://doi.org/10.1063/5.0011479]

\bibitem{sdi_ave}
M.~Avesani, H.~Tebyanian, P.~Villoresi, and G.~Vallone,
``Semi-Device-Independent Heterodyne-Based Quantum Random-Number Generator,''
Phys. Rev. Applied \textbf{15}, 034034 (2021)
doi:10.1103/PhysRevApplied.15.034034
[link.aps.org/doi/10.1103/PhysRevApplied.15.034034]

\bibitem{DIBELL}
Yang Liu, Qi Zhao, Ming-Han Li, Jian-Yu Guan, Yanbao Zhang, Bing
Bai, Weijun Zhang, Wen-Zhao Liu, Cheng Wu, Xiao Yuan, Hao Li, W.
J. Munro, Zhen Wang, Lixing You, Jun Zhang, Xiongfeng Ma, Jingyun
Fan, Qiang Zhang, and Jian-Wei Pan,
  \emph{Device-independent quantum random-number generation},
  Nature, Vol. 562, No. 7728, pp. 548--551, Oct 2018.
  \href{http://www.nature.com/articles/s41586-018-0559-3}{doi:10.1038/s41586-018-0559-3}

\bibitem{SDI_Avesani2022}
M.~Avesani, H.~Tebyanian, P.~Villoresi, and G.~Vallone,
``Unbounded randomness from uncharacterized sources,''
Communications Physics \textbf{5}, 273 (2022)
doi:10.1038/s42005-022-01038-3
[url: https://doi.org/10.1038/s42005-022-01038-3]

\bibitem{DI_me}
G.~Foletto, M.~Padovan, M.~Avesani, H.~Tebyanian, P.~Villoresi, and G.~Vallone,
``Experimental test of sequential weak measurements for certified quantum randomness extraction,''
Phys. Rev. A \textbf{103}, 062206 (2021)
doi:10.1103/PhysRevA.103.062206
[url: https://link.aps.org/doi/10.1103/PhysRevA.103.062206]

\bibitem{semi-di-new}
H. Tebyanian, M. Zahidy, R. Müller, et al., ``Generalized time-bin quantum random number generator with uncharacterized devices,'' \textit{EPJ Quantum Technol.}, vol. 11, no. 15, 2024. [Online]. Available: \url{https://doi.org/10.1140/epjqt/s40507-024-00227-z}

\bibitem{QKD21}
M.~Avesani, L.~Calderaro, G.~Foletto, C.~Agnesi, F.~Picciariello, F. B. L.~Santagiustina, A.~Scriminich, A.~Stanco, F.~Vedovato, M.~Zahidy, G.~Vallone, and P.~Villoresi,
``Resource-effective quantum key distribution: a field trial in Padua city center,''
Opt. Lett. \textbf{46}, 2848--2851 (2021)
doi:10.1364/OL.422890
[url:http://opg.optica.org/ol/abstract.cfm?URI=ol-46-12-2848]

\bibitem{QKD}
Y.~Ding, D.~Bacco, K.~Dalgaard, X.~Cai, X.~Zhou, K.~Rottwitt, and L. K.~OxenlÃ¸we,
``High-dimensional quantum key distribution based on multicore fiber using silicon photonic integrated circuits,''
npj Quantum Information \textbf{6}, 1-7 (2020)
doi:10.1038/s41534-020-0259-9
[url:https://doi.org/10.1038/s41534-020-0259-9]

\bibitem{LHV}
John F. Clauser, Michael A. Horne, Abner Shimony, and Richard A.
Holt,
  \emph{Proposed Experiment to Test Local Hidden-Variable Theories},
  Phys. Rev. Lett., Vol. 23, No. 15, pp. 880--884, Oct 1969.
  \href{https://link.aps.org/doi/10.1103/PhysRevLett.23.880}{doi:10.1103/PhysRevLett.23.880}

\bibitem{BOX}
P.~Janotta, ``Generalizations of Boxworld,'' \textit{Electronic Proceedings in Theoretical Computer Science}, vol. 95, pp. 183--192, Oct. 2012. doi:10.4204/eptcs.95.13


\end{thebibliography}
\end{document}